\documentstyle[preprint,aps]{revtex}

\begin{document}

\draft


\title{Notes on Spinning AdS$_{3}$ Black Hole Solution}

\author{Hongsu Kim}

\address{Department of Physics\\
Ewha Women's University, Seoul 120-750, KOREA}

\date{November, 1996}

\maketitle

\begin{abstract}
By applying Newman's method, the AdS$_{3}$ rotating black hole solution
is ``derived" from the nonrotating black hole solution of 
Ba$\tilde{\rm{n}}$ados, Teitelboim, and Zanelli (BTZ). 
The rotating BTZ solution derived in this fashion is given in
``Boyer-Lindquist-type'' coordinates whereas the form of the solution
originally given by BTZ is given in a kind of an ``unfamiliar'' 
coordinates which are related to each other by a 
transformation of time coordinate alone. 
The relative physical meaning between these two time coordinates is
carefully studied by evaluating angular momentum per unit mass, 
angular velocity, surface gravity and area of the event horizon in two 
alternative coordinates respectively.
The result of this study leads us to the conclusion that the  BTZ time 
coordinate must be the time coordinate of an observer who rotates around
the axis of the spinning hole in opposite direction to that  of the hole 
outside its static limit.

\end{abstract}


\narrowtext




{\bf I. Introduction}
\\
It had long been thought that black hole solutions cannot exit in 3-dim. since
there is no local gravitational attraction and hence no mechanism to confine
large densities of matter. It was, therefore, quite a surprise when 
Ba$\tilde{\rm{n}}$ados,
Teitelboim, and Zanelli (BTZ) [1]  have recently constructed the 
Anti-de Sitter (AdS$_{3}$)
spacetime solution to the Einstein equations in 3-dim. that can be interpreted
as
a black hole solution. They included the negative cosmological constant in the
3-dim. vacuum Einstein theory and then found both the rotating and nonrotating
black hole solutions. In the mean time, a curious relationship between the
nonrotating and the rotating spacetime solutions of Einstein theory in 4-dim. 
also has been long known. Newman et al. [2] discovered long ago that one can 
``derive"
Kerr solution from the Schwarzschild solution in vacuum Einstein theory and 
Kerr-Newman solution from the Reissner-Nordstr$\ddot{\rm{o}}$m solution in 
Einstein-Maxwell
theory via the ``complex coordinate transformation" scheme acting on metrics
written in terms of null tetrad of basis vectors. In the present work, we 
attempt the
same derivation but in this time in 3-dim. spacetime. Namely, we see if the 
rotating
version of BTZ black hole solution can indeed be ``derived" from its 
nonrotating
counterpart via Newman's method. And in doing so, our philosophy is that the
3-dim. situation can be thought of as, say, the $\theta = \pi/2$ - slice of the
4-dim. one (where $\theta$ denotes the polar angle). 
Interestingly enough, we do end up with the rotating version of BTZ
black hole solution but in a different coordinate system from the one 
originally
employed in BTZ's solution ansatz. And as we shall see shortly, it turns out
that the rotating BTZ solution derived here in this work is given in
``Boyer-Lindquist-type'' [3] coordinates whereas the original form of the 
solution given by BTZ is given in a kind of an ``unusual'' coordinates which 
are related to
the ``familiar'' Boyer-Lidquist-type coordinates by a transformation of time
coordinate alone. We can easily understand the reason for this result as 
follows ; much as the Kerr solution ``derived'' from the Schwarzschild
solution by Newman's complex coordinate transformation method naturally
comes in Boyer-Lindquist coordinates [3] (of course via a coordinate 
transformation from Kerr coordinates [3]), the rotating BTZ black hole solution
derived from its nonrotating counterpart in this manner comes in
Boyer-Lindquist-type coordinates as well. And then one can realize that by
performing a transformation from the Boyer-Lindquist-type time coordinate
$t$ to a ``new'' time coordinate $\tilde{t}$ following the transformation law,
$\tilde{t} = t - a \phi$ (where $a$ is proportional to the angular momentum 
per unit mass and $\phi$ denotes the azimuthal angle), 
our ``derived'' rotating BTZ black hole solution can indeed be put
in the form originally given by BTZ.
As expected, the rotating BTZ black hole solution given in Boyer-Lindquist-type
coordinates takes on the structure which resembles that of Kerr solution
more closely than it is in BTZ's rather unusual time coordinate. 
However, one can readily realize that it is the BTZ's time coordinate 
$\tilde{t}$ that is the usual Killing time coordinate, not the Boyer-Lindquist-
type one.
And in order to investigate the relative physical meaning between these two
time coodinates, quantities like angular momentum, angular velocity,
surface gravity and area of the event horizon are evaluated in two 
alternative coordinates. The analysis of this observation on the behavior
of the black hole solution with two different choices of coordinates will
be presented later on in the discussion. 
\\
{\bf II. Derivation of the AdS$_{3}$ rotating black hole solution}
\\
As mentioned earlier, Newman et al. [2] discovered curious ``derivations" of 
stationary,
axisymmetric metric solutions from static, spherically-symmetric solutions
in 4-dim. Einstein theory. To attempt the similar work in 3-dim. spacetime, we
start with the nonrotating version of BTZ black hole solution in 3-dim. as a 
``seed" 
solution to construct its rotating counterpart. The nonrotating BTZ black hole
solution written in Schwarzschild-type coordinates $(t, r, \phi)$ is given by
\begin{eqnarray}
ds^2 = (-M + {r^2\over l^2})dt^2 - (-M + {r^2\over l^2})^{-1}dr^2 - r^2d\phi^2
\end{eqnarray}
where $l$ is related to the negative cosmological constant by $l^{-2} = 
-\Lambda$
and $M$ is an integration constant that can be identified with the ADM mass 
of the black hole. Now in order to ``derive" a rotating black hole solution 
applying the
complex coordinate transformation scheme of Newman et al., we begin by assuming
that this 3-dim. black hole geometry is a $\theta = \pi/2$ - slice of a static,
spherically-symmetric 4-dim. geometry given by
\begin{eqnarray}
ds^2_{4} = \lambda^{2}(r)dt^2 - \lambda^{-2}(r)dr^2 - r^2 (d\theta^2 + 
\sin^2 \theta d\phi^2)
\end{eqnarray} 
with $\lambda^{2}(r) = (-M+r^{2}/l^{2})$. The essential reason for this 
``dimensional continuation" is to introduce the null tetrad system of vectors
on which Newman's complex coordinate transformation method is technically 
based.
We do not, however, ask nor demand that this 4-dim. geometry be an explicit
solution of Einstein equation in 4-dim. as well. We just demand that only its
$\theta = \pi/2$ - slice be a solution of Einstein equation in 3-dim.
Remarkably, then, upon the series of operations ; dimensional continuation 
$\rightarrow $ Newman's derivation method $\rightarrow $ dimensional reduction
by setting $\theta = \pi/2$, we end up with a legitimate rotating black hole
solution to 3-dim. Einstein equation as we shall see shortly.
\\
Consider now the transformation to the Eddington-Finkelstein-type retarded 
null coordinates $(u, r, \phi)$ defined by $u = t - r_{\ast}$, with 
$r_{\ast} = \int dr (g_{rr}/-g_{tt})^{1/2}$. In terms of these null 
coordinates, the nonrotating BTZ black hole metric takes the form
\begin{eqnarray}
ds^2_{4} = \lambda^{2}(r)du^2 + 2du dr - r^2 (d\theta^2 +
\sin^2 \theta d\phi^2).
\end{eqnarray}
Into this 4-dim. Riemannian space we next introduce a tetrad system of vectors 
$l_{\mu}$,
$n_{\mu}$, $m_{\mu}$ and $\bar{m}_{\mu}$ (with $\bar{m}_{\mu}$ being the 
complex 
conjugate of $m_{\mu}$) satisfying the following orthogonality property
$l_{\mu}n^{\mu} = - m_{\mu}\bar{m}^{\mu} = 1$ with all other scalar products
vanishing. In terms of this null tetrad of basis vectors, the spacetime 
metric is written as $g^{\mu \nu} = l^{\mu}n^{\nu} + n^{\mu}l^{\nu} - 
m^{\mu}\bar{m}^{\nu} - \bar{m}^{\mu}m^{\nu}$. Then now one can obtain the 
contravariant components of
the metric and the null tetrad vectors from the covariant components of the 
metric given in eq.(3) as
\begin{eqnarray}
&g^{00}&=0, ~~~g^{11}=-\lambda^2(r), ~~~g^{10}=1, \\
&g^{22}&=-{1\over r^2}, ~~~g^{33}=-{1\over r^2 \sin^2 \theta} \nonumber
\end{eqnarray}  
and
\begin{eqnarray}
&l^{\mu}& = \delta^{\mu}_{1}, ~~~n^{\mu} = \delta^{\mu}_{0} - {1\over 2}
\lambda^2(r)\delta^{\mu}_{1}, \nonumber \\
&m^{\mu}& = {1\over \sqrt{2}r}(\delta^{\mu}_{2} + {i\over \sin \theta}
\delta^{\mu}_{3}), \\
&\bar{m}^{\mu}& = {1\over \sqrt{2}r}(\delta^{\mu}_{2} - {i\over \sin \theta}
\delta^{\mu}_{3}) \nonumber
\end{eqnarray}
respectively. Now the radial coordinate $r$ is allowed to take complex values
and the tetrad is rewritten in the form
\begin{eqnarray}
&l^{\mu}& = \delta^{\mu}_{1}, ~~~n^{\mu} = \delta^{\mu}_{0} - {1\over 2}
(-M+{r\bar{r}\over l^2})\delta^{\mu}_{1}, \nonumber \\
&m^{\mu}& = {1\over \sqrt{2}\bar{r}}(\delta^{\mu}_{2} + {i\over \sin \theta}
\delta^{\mu}_{3}), \\
&\bar{m}^{\mu}& = {1\over \sqrt{2}r}(\delta^{\mu}_{2} - {i\over \sin \theta}
\delta^{\mu}_{3}) \nonumber
\end{eqnarray}
with $\bar{r}$ being the complex conjugate of $r$ (note that part of the 
algorithm
is to keep $l^{\mu}$ and $n^{\mu}$ real and $m^{\mu}$ and $\bar{m}^{\mu}$ the
complex conjugate of each other). We now formally perform the ``complex 
coordinate transformation"
\begin{eqnarray}
r' &=& r + i a \cos \theta,  ~~~\theta' = \theta, \\
u' &=& u - i a \cos \theta,  ~~~\phi' = \phi \nonumber
\end{eqnarray}
on tetrad vectors $l^{\mu}$, $n^{\mu}$ and $m^{\mu}$ ($\bar{m}'^{\mu}$ is, 
as stated, defined as the complex conjugate of ${m}'^{\mu}$). 
If one now allows $r'$ and $u'$ to be real, we obtain the following tetrad
\begin{eqnarray}
&l'^{\mu}& = \delta^{\mu}_{1}, \\
&n'^{\mu}& = \delta^{\mu}_{0} - {1\over 2}
[-M+{(r'^2+a^2\cos^2 \theta) \over l^2}]\delta^{\mu}_{1}, \nonumber \\
&m'^{\mu}& = {1\over \sqrt{2}(r'+ia\cos \theta)}[ia\sin \theta 
(\delta^{\mu}_{0} - \delta^{\mu}_{1}) +
\delta^{\mu}_{2} + {i\over \sin \theta}\delta^{\mu}_{3}] \nonumber
\end{eqnarray}
from which one can readily read off the contravariant components of the 
metric and 
then obtain the covariant components by inversion as (henceforth we drop the
``prime")
\begin{eqnarray}
ds^2_{4} &=& (-M+{\Sigma \over l^2})du^2 + 2a\sin^2 \theta [1-(-M+{\Sigma 
\over l^2})] du d\phi + 2 du dr \nonumber \\
&-& 2a\sin^2 \theta dr d\phi - \Sigma d\theta^2 -
[r^2+a^2+a^2\sin^2 \theta \{1-(-M+{\Sigma \over l^2})\}]\sin^2 \theta d\phi^2 
\end{eqnarray}
where $\Sigma \equiv (r^2+a^2\cos^2 \theta)$.
Now at this stage, considering that the geometry
in 3-dim. can be thought of as the $\theta = \pi/2$-slice of the full, 4-dim. 
one,
we simply set $\theta = \pi/2$ in the metric above to arrive at the rotating
black hole metric in 3-dim. given by
\begin{eqnarray}
ds^2 = (-M+{r^2\over l^2})(du - ad\phi)^2 
+ 2(du - ad\phi)(dr + ad\phi) - r^2 d\phi^2.
\end{eqnarray} 
Also note that the metric above we ``derived" via Newman's complex coordinate
transformation method is given in terms of Kerr-type coordinates [3]
$(u, r, \phi)$
which can be thought of as the generalization of the retarded null coordinates.
Thus one might want to further transform it into the one written in
Boyer-Lindquist-type coordinates [3]
$(t, r, \hat{\phi})$ that can be viewed as the generalization of the 
Schwarzschild coordinates. This can be achieved via the transformation
\begin{eqnarray}
dt = du + {(r^2+a^2)\over \Delta}dr, 
~~~d\hat{\phi} = d\phi + {a\over \Delta}dr 
\end{eqnarray}
where $\Delta \equiv r^2(-M+r^2/ l^2)+a^2$.
Finally, the rotating AdS$_{3}$ black hole solution  given in Boyer-Lindquist-
type coordinates is given by (henceforth we drop ``hat" on $\phi$ coordinate)
\begin{eqnarray}
ds^2 &=& (-M+{r^2\over l^2})dt^2 + 2a[1-(-M+{r^2\over l^2})]dt d\phi \\
&-& [r^2+a^2+a^2\{1-(-M+{r^2\over l^2})\}]d\phi^2 - {r^2\over \Delta}dr^2. \nonumber
\end{eqnarray}
And it is straightforward to check that the ``derived''
rotating black hole solution given in eq.(12) does satisfy the AdS$_{3}$
Einstein
equation, $R_{\mu \nu} - {1\over 2}g_{\mu \nu}R  - l^{-2}g_{\mu \nu} = 0$.\\
It is a little puzzling, however, that the derived rotating black hole 
solution of ours in eq.(12) above does not appear exactly the same as the one 
originally constructed by BTZ.
Nonetheless, both our solution above and the one originally
obtained by BTZ (written in our notation convention in which $a = J/2$ with 
$J$ appearing in original BTZ's work [1])
\begin{eqnarray}
ds^2 = (-M+{r^2\over l^2})d\tilde{t}^2 + 2ad\tilde{t}d\phi - r^2d\phi^2 - 
{r^2\over \Delta}dr^2
\end{eqnarray}
correctly reduce, in the vanishing angular momentum limit $a\rightarrow 0$, to
the static, spherically-symmetric black hole solution which was the starting
point of our solution construction. 
Therefore, presumably the two rotating black hole solutions must be related to
each other by a coordinate transformation.
Indeed, one can check straightforwardly that the two are related by the 
transformation of the time coordinate alone
\begin{eqnarray}
\tilde{t} = t - a\phi.
\end{eqnarray} 
Therefore the two metrics given in eqs.(12) and (13) represent one and the same
AdS$_{3}$ black hole solution modulo gauge transformation.
Notice, however, that the rotating BTZ black hole solution given in the
Boyer-Lindquist-type time coordinate in eq.(12) resembles the structure
of Kerr solution in 4-dim. more closely than that given in the BTZ time
coordinate in eq.(13).
It seems now that we are left with the question ; between $t$ and $\tilde{t}$,
which one is the usual Killing time coordinate ?
It appears that it is $\tilde{t}$, the BTZ time coordinate that is
the usual Killing time coordinate, {\it not} $t$, the Boyer-Lindquist-type 
time coordinate. And this conclusion is based on the following observation.
Note that both in Boyer-Lindquist-type coordinates $(t, ~r, ~\phi)$ and in
BTZ coordinates $(\tilde{t}, ~r, ~\phi)$, generally $\mid g_{t\phi} \mid$
$(\mid g_{\tilde{t}\phi} \mid)$ represents {\it angular momentum per unit 
mass} as observed by an accelerating observer at some fixed $r$. 
(We stress here that the quantity we are about to introduce is the
angular momentum {\it per unit mass} at some point $r$. 
Generally, it should be
distinguished from the total angular momentum of the spacetime measured
in the asymptotic region, 
$\tilde{J} = (16\pi )^{-1}\int_{S}\epsilon_{\mu\nu\alpha\beta}
\nabla^{\alpha}\psi^{\beta}$ in the notation convention of ref.[4]
with $S$ being a large sphere in the asymptotic region and
$\psi^{\mu} = (\partial/\partial \phi)^{\mu}$ being the rotational
Killing field. Thus in these definitions, angular momentum {\it per
unit mass} may change under a coordinate transformation although
the total angular momentum $\tilde{J}$ remains coordinate independent.)
To see this quickly, one only needs to write the rotating black hole metric
in ADM's (2+1)-split form
\begin{eqnarray}
ds^2 = N^2(r)dt^2 - f^{-2}(r)dr^2 - R^2(r)[N^{\phi}(r)dt + d\phi]^2.
\end{eqnarray}
Then in Boyer-Lindquist-type time coordinate $t$, metric components
correspond to
\begin{eqnarray}
f^{2}(r) &=& (-M + {r^2 \over l^2} + {a^2 \over r^2}), \nonumber \\
R^2(r) &=& [r^2 + a^2 +a^2 \{1 - (-M + {r^2 \over l^2})\}], \\
N^2(r) &=& (-M + {r^2 \over l^2}) + R^{-2}(r) a^2 \{1 - (-M + {r^2 \over
l^2})\}^2, \nonumber \\
N^{\phi}(r) &=& -R^{-2}(r) a \{1 - (-M + {r^2 \over l^2})\} \nonumber
\end{eqnarray}
whereas in BTZ time coordinate $\tilde{t}$, they correspond to
\begin{eqnarray}
N^2(r) &=& f^{2}(r) = (-M + {r^2 \over l^2} + {a^2 \over r^2}), \nonumber\\
N^{\phi}(r) &=& -{a \over r^2}, ~~~~R^2(r) = r^2. \nonumber
\end{eqnarray}
Now it is apparent in this ADM's space-plus-time split form of the
metric given in eq.(15) that the shift function $N^{\phi}(r)$ corresponds
to the angular velocity $\mid N^{\phi}(r) \mid = \Omega(r)$ and the
quantity $R^2(r) \mid N^{\phi}(r) \mid = J$ can be identified with
the angular momentum per unit mass as observed at some point $r$ outside
the rotating hole. \\
Thus from $-R^2(r)N^{\phi}(r) = g_{t\phi}$, the angular momentum 
per unit mass in
Boyer-Lindquist-type time coordinate and in BTZ time coordinate are
given respectively by
\begin{eqnarray}
J &=& \mid g_{t\phi} \mid = a [1 - (-M + {r^2 \over l^2})], \nonumber \\
J^{BTZ} &=& \mid g_{\tilde{t} \phi} \mid = a 
\end{eqnarray}
and thus in particular, the angular momentum per unit mass at the event
horizon is given respectively by
\begin{eqnarray}
J_{H} &=& \mid g_{t\phi} (r_{+}) \mid = a {(r^2_{+} + a^2)\over r^2_{+}}, 
\nonumber \\
J^{BTZ}_{H} &=& \mid g_{\tilde{t}\phi} (r_{+}) \mid = a
\end{eqnarray}
indicating that in BTZ time coordinate the spinning hole appears to
have smaller angular momentum per unit mass at the 
event horizon (and generally
everywhere outside the hole) than it does in Boyer-Lindquist-type
time coordinate. Now from eq.(17), it is manifest that the angular 
momentum per unit mass given in BTZ time coordinate $\tilde{t}$ is
finite and constant all over the hypersurface whereas that given
in Boyer-Lindquist-type time coordinate $t$ grows indefinitely
as $r \rightarrow \infty$.
This can be attributed to the fact that the coordinate $\tilde{t}$
BTZ used is defined asymptotically using the asymptotic symmetries
and hence approaches the AdS time at spatial infinity [1].
Therefore we can conclude that it is the BTZ time coordinate
$\tilde{t}$ which is the usual Killing time. This is certainly
in contrast to what happens in the familiar Kerr black hole
geometry in 4-dim. where the usual Boyer-Lindquist time coordinate
is the Killing time coordinate. And it seems that this discrepancy 
comes from the fact that the 3-dim. BTZ black hole is not
asymptotically flat but asymptotically anti-de Sitter whereas
the 4-dim. Kerr black hole is asymptotically flat.
The final question one might want to ask and answer could then be ; what is
the relative physical meaning of these two time coordinates 
$t$ and $\tilde{t}$ ?
To get a quick answer to this question, we go back and look at the
coordinate transformation law given in eq.(14) relating the two
time coordinates $t$ and $\tilde{t}$. Namely, taking the dual of
the transformation law
$\delta \tilde{t} = \delta t - a \delta \phi$, we get
\begin{eqnarray}
({\partial \over \partial \tilde{t}})^{\mu} &=&
({\partial \over \partial t})^{\mu} - {1\over a}
({\partial \over \partial \phi})^{\mu} \nonumber \\
{\rm or} \nonumber \\
\tilde{\xi}^{\mu} &=& \xi^{\mu} - {1\over a}\psi^{\mu} \nonumber
\end{eqnarray}
where $\xi^{\mu} = (\partial /\partial t)^{\mu}$ and
$\psi^{\mu} = (\partial /\partial \phi)^{\mu}$ denote Killing 
fields corresponding to the time translational and the rotational
isometries of the spinning black hole spacetime respectively and
$\tilde{\xi}^{\mu} = (\partial /\partial \tilde{t})^{\mu}$ denotes
the Killing field associated with the isometry of the hole's metric
under the BTZ time translation. Now this expression for the BTZ
time translational Killing field $\tilde{\xi}^{\mu}$ implies that
in BTZ time $\tilde{t}$, the time translational generator is
given by the linear combination of the Boyer-Lindquist-type time 
translational generator and the rotational generator. In plain
English, this means that in BTZ time coordinate, the action of
time translation consists of the action of Boyer-Lindquist-type time
translation and the action of rotation in opposite direction to
$a$, i.e., to the rotation direction of the hole. Thus the BTZ
time coordinate $\tilde{t}$ can be interpreted as the coordinate,
say, of a frame which rotates around the axis of the spinning
BTZ black hole in opposite direction to that of the hole.
Furthermore, by considering the angular velocity, the angular
momentum per unit mass, the surface gravity and the area of 
the event horizon of the hole both in Boyer-Lindquist-type
time and in BTZ time coordinate and then comparing them, one
can explore the relative physical meaning between the two
time coordinates in a more comprehensive manner.
And to do so, we need to compare the causal structures
and the black hole thermodynamics investigated in the two alternative
time coordinates.  (The reason for doing this is, as we shall see, 
aside from the value of studying causal structure and
thermodynamics themselves, they also will provide us with very 
clear and natural interpretation of the two alternative time 
coordinates $t$ and $\tilde{t}$.)
Since they have already been studied in the BTZ time 
coordinate $\tilde{t}$ by BTZ in their original works, here in this work we
consider them in the Boyer-Lindquist-type time coordinate.
As we shall see shortly, the causal structure and the global topology remain
unaffected under the coordinate change $\tilde{t} = t - a\phi$ as they shoud.
An important point, however, is that quantities like angular momentum per
unit mass,
angular velocity and the surface gravity at the event horizon are obtained 
{\it differently} in the two
alternative time coordinates. And it is precisely this point that will 
enable us to realize the relative physical meaning of the two time 
coordinates as we shall see in the following section.
\\
{\bf III. Causal structure and black hole thermodynamics}
\\ 
Now consider the causal structure [4] of the rotating BTZ black hole 
solution in the Boyer-Lindquist-type time coordinates.  As we
shall see shortly, it turns out that its causal structure is almost the same as
that in the BTZ time coordinate and their global structures are exactly the 
same
as is manifest in their Carter-Penrose diagrams (we shall not provide the 
diagrams here since they can be found in the literature).
First we start with the curvature singularity. Inspection of the behavior 
of typical
curvature invariant shows that it is everywhere regular including $r = 0$. 
In fact, this was expected since generally the (anti-)de Sitter spacetime is a
maximally-symmetric space with a constant curvature.
Next we consider the event horizons.
As mentioned earlier, our rotating AdS$_{3}$ black hole 
solution
is stationary and axisymmetric and thus possesses two Killing fields
$\xi^{\mu} = (\partial / \partial t)^{\mu}$ and 
$\psi^{\mu} = (\partial / \partial \phi)^{\mu}$ correspondingly.
And it is their linear combination
$\chi^{\mu} = \xi^{\mu} + \Omega_{H}\psi^{\mu}$ which is normal to the Killing
horizons of the black hole spacetime. In fact, normaly this is the defining 
equation
of the angular velocity of the event horizon, $\Omega_{H}$. Now since the 
Killing
horizon is defined to be a surface on which the Killing field $\chi^{\mu}$ 
becomes
null, in order to find the event horizon we should look for zeros of
$\chi^{\mu}\chi_{\mu} = 0$. A straightforward calculation shows that the 
Killing
field $\chi^{\mu}$ becomes null at points where
$\Delta = r^2(-M + r^2/l^2) + a^2 = 0$. Thus we have regular inner and outer
horizons at
\begin{eqnarray}
r_{\pm} = l [{M\over 2}\{1 \pm \sqrt{1 - ({2a\over Ml})^2}\}]^{1/2}
\end{eqnarray}
(i.e., $M = (r^2_{+}+r^2_{-})/l^2$ and $a = r_{+}r_{-}/l$)
with $r_{+}$being the black hole event horizon provided $|a| \leq Ml/2$.
Note that these coordinate singularities are absent in Kerr-type null 
coordinates
although manifest in Boyer-Lindquist-type coordinates. And the angular velocity
of this event horizon is given by
\begin{eqnarray}
\Omega_{H} = -{g_{t\phi}\over g_{\phi \phi}}\vert_{r_{+}} = {a\over 
(r^2_{+} + a^2)}.
\end{eqnarray}
Also note that rotating BTZ black hole solution given in Boyer-Lindquist-type 
time coordinate develops inner and outer horizons
exactly at the same locations as those in BTZ time coordinate. However, 
there is an
important difference ; the angular velocity of the event horizon evaluated
in Boyer-Lindquist-type time coordinate,
$\Omega_{H} = a/(r^2_{+}+a^2)$ is smaller than that in BTZ time coordinate
\begin{eqnarray}
\Omega^{BTZ}_{H} =  -{g_{\tilde{t}\phi}\over g_{\phi \phi}}\vert_{r_{+}}
= {a \over r^2_{+}}
\end{eqnarray}
and thus the hole appears to {\it spin more slowly} in Boyer-Lindquist-type
time coordinate.
Finally, we consider the ``static limit" which is the outer boundary of the
ergoregion. Note that the time-translational Killing field with the norm
$\xi^{\mu}\xi_{\mu} = g_{tt} = (-M + r^2/l^2)$ becomes spacelike, null and then
timelike as $r$ increases. Particularly note that the region in which 
$\xi^{\mu}$
stays spacelike extends outside the black hole's event horizon. This region is
usually called ``ergoregion" and its outer boundary on which $\xi^{\mu}$ 
becomes
null is called ``static limit" since inside of which no observer can possibly
remain static. Thus in order to find the location of this static limit, we look
for a zero of $\xi^{\mu}\xi_{\mu} = 0$. And it turns out that the static limit
of the rotating BTZ black hole solution in Boyer-Lindquist-type coordinate 
occurs at $r_{s} = \sqrt{M}l > r_{+}$ again
exactly at the same location as that in BTZ time coordinate.
\\
Finally, we discuss the thermodynamics of the rotating black hole solution
given in Boyer-Lindquist-type time coordinate.
Since the practical study of black hole thermodynamics [4] begins and ends 
with the
temperature and entropy of the hole, we shall attempt to compute them.
We first begin with the Hawking temperature $T_{H}$ measured by an
observer in the asymptotic region. The Hawking temperture $T_{H}$ is 
related to the surface gravity $\kappa$ of the black hole by the relation
$T_{H} = \kappa/2\pi$ which is supposed to hold for any stationary black 
hole [6].
Thus our task is simply to calculate the surface gravity of the hole.
In physical terms, the surface gravity $\kappa$ is the force that must 
be exerted
to hold a unit test mass at the horizon and it is given in a simple formula 
as [4]
\begin{eqnarray}
\kappa^2 = -{1\over 2}(\nabla^{\mu}\chi^{\nu})(\nabla_{\mu}\chi_{\nu})
\end{eqnarray}
where $\chi^{\mu}$ is as given earlier and the evaluation on the event horizon
$r = r_{+}$ is assumed. A direct computation then gives the Hawking temperature
as 
\begin{eqnarray}
T_{H} = {\kappa\over 2\pi} = {1\over 2\pi l^2}{r_{+}(r^2_{+}-r^2_{-})\over
(r^2_{+}+a^2)} = {1\over 2\pi r_{+}}({r^2_{+}-r^2_{-}\over r^2_{-}+l^2}) 
\end{eqnarray}
where we used $M = (r^2_{+}+r^2_{-})/l^2$ and $a = r_{+}r_{-}/l$.
Recall that the Hawking temperature
(or surface gravity) of the rotating BTZ solution in BTZ time coordinate is
given by 
$T^{BTZ}_{H} = (r^2_{+} - r^2_{-})/2\pi l^2 r_{+}$.
Next, we turn to the computation of the black hole's entropy. Generally, the
black hole entropy (i.e., the semiclassical entropy, not the fine-grained,
quantum statistical one) can be evaluated in three ways : first, following
Bekenstein-Hawking proposal [5, 6], one can argue {\it a priori} that 
the entropy of
a black hole must be proportional to the surface area of its event horizon
$(S = A/4)$. Alternatively, knowing the Hawking temperature $T_{H}$ and the
chemical potential (i.e., $\Omega_{H}$ for the conserved angular momentum
$J_{H}$), one may integrate the 1st law of black hole thermodynamics 
$T_{H}dS = dM - \Omega_{H} dJ_{H} +$ (variation of gravitational source
terms), to obtain the entropy S.
Lastly, according to Gibbons and Hawking [7], thermodynamic functions 
including
the black hole entropy can be computed directly from the saddle point
approximation to the gravitational partition function
$S = \ln Z + \beta (H - \Omega_{H} J_{H}) \simeq -I[g^{c}] + 
(M - \Omega_{H} J_{H})/T_{H}$
where $I[g^{c}]$ is the Euclidean action evaluated at the saddle point $g^{c}$,
i.e., the classical metric solution and $Z$ is the partition function.
Although the second method turns out to be inappropriate for asymptotically
non-flat cases like this AdS$_{3}$ black hole case, 
the first and the third methods, of course, agree to yield 
the same expression for the entropy as
\begin{eqnarray}
S = {1\over 4}A = {\pi (r^2_{+} + a^2)\over 2r_{+}} = {\pi r_{+}\over 2l^2}
(r^2_{-} + l^2)
\end{eqnarray}
where again we used $a = r_{+}r_{-}/l$. For the sake of comparison, recall that
the entropy of the rotating BTZ solution in BTZ time coordinate is given by
$S^{BTZ} = \pi r_{+}/2$. \\
Now we summarize what we have learned concerning the causal 
structures and the
black hole thermodynamics investigated in the two alternative time coordinates.
It turned out that whichever time coordinate, the Boyer-Lindquist-type one $t$
or the BTZ one $\tilde{t}$, one may take, one ends up with the same causal
structure and the global topology except for the fact,
\begin{eqnarray}
J_{H} &=& ({r^2_{+}+a^2 \over r^2_{+}}) J^{BTZ}_{H} > J^{BTZ}_{H}, \nonumber\\
\Omega_{H} &=& ({r^2_{+}\over r^2_{+}+a^2}) \Omega^{BTZ}_{H} < 
\Omega^{BTZ}_{H},  \\
T_{H} &=& ({r^2_{+}\over r^2_{+}+a^2}) T^{BTZ}_{H} < 
 T^{BTZ}_{H}, \nonumber \\
S &=& ({r^2_{+}+a^2 \over r^2_{+}}) S^{BTZ} > 
 S^{BTZ} \nonumber
\end{eqnarray}
which are obtained from eqs.(18), (20), (21), (23) and (24) and
where quantities with the superscript ``BTZ'' denote the ones computed in
BTZ time coordinate $\tilde{t}$, whereas quantities without superscript
denote the ones computed in Boyer-Lindquist-type time coordinate $t$. \\
These results indicate that in BTZ time coordinate $\tilde{t}$, the
rotating BTZ black hole solution has smaller angular momentum 
per unit mass yet greater
angular velocity, greater surface gravity and smaller area of the event
horizon than it does in the Boyer-Lindquist-type time coordinate $t$.
Particularly here, ``possessing smaller angular momentum per 
unit mass while greater angular velocity'' may first look erroneous but 
if one really looks into
the details one can realize that it is no nonsense since it arises from
``same coordinate distance but different proper distances'' to the horizon
in the two time coordinates $t$ and $\tilde{t}$. \\
Now, first from the greater angular velocity, we are led to the conclusion
that the BTZ time coordinate $\tilde{t}$ must be the one, say, of an 
observer
who {\it rotates around the axis of the spinning hole in opposite direction
to the rotation direction of the hole outside its static limit}.
This interpretation, then, readily explains the smaller length of the
horizon (horizon area becomes {\it length} in 3-dim.) as being the
Lorentz length contraction. Next, smaller angular momentum per 
unit mass and greater
surface gravity can be attributed to the fact that as the angular
momentum per unit mass decreases when transforming from the 
Boyer-Lindquist-type 
to BTZ time coordinate, the surface gravity is expected to increase 
due to  an effect of centrifugal force. 
\\
{\bf IV. Discussions}
\\
To conclude, here in this work we attempted to apply Newman's method
of generating a spinning black hole solution from a static black hole
solution in 4-dim. Einstein theory to the 3-dim. situation.
More concretely, we employed the algorithm ; dimensional continuation
$\rightarrow $ Newman's derivation method $\rightarrow $ dimensional
reduction by setting $\theta = \pi/2$ to successfully obtain a legitimate
rotating AdS$_{3}$ black hole solution.
And as a consequence of this study, we discovered that there are two
alternative time coordinates in describing
the rotating AdS$_{3}$ black hole solution one can select from
to investigate its various physical contents.
Thus let us elaborate on the complementary roles played by the 
two alternative time coordinates. 
First, note that the two time coordinates, that of  Boyer-Lindquist-type,
$t$ and that of BTZ, $\tilde{t}$ coincide for nonrotating case
$(a = 0)$ and become different only for rotating case $(a \neq 0)$
as one can see in their relation, $\tilde{t} = t - a\phi$.
Next, the Boyer-Lindquist-type time coordinate
$t$ is the usual time coordinate with which one can obtain the
rotating black hole's characteristics such as the angular velocity of 
the event horizon or the surface gravity as measured by an outside
observer who is ``static'' with respect to, say, a distant body
just as it is the case with Boyer-Lindquist time coordinate in
Kerr black hole spacetime in 4-dim.
The BTZ time coordinate $\tilde{t}$, on the other hand, may look
``unusual''
in that it can be identified with the time coordinate of a non-static
observer who rotates in opposite direction to that of the spinning hole.
It is, however, this BTZ time coordinate, not the familiar
Boyer-Lindquist-type time coordinate, which is the right Killing
time coordinate in terms of which the total mass and particularly
the  angular momentum can be well-defined in the asymptotic
region of this asymptotically anti-de Sitter spacetime.
Also it has been well-studied in the literature [1] that this BTZ time 
coordinate
is particularly advantageous in exploring the global structure of the
rotating BTZ black hole since it allows one to transform to Kruskal-type 
coordinates and eventually allows one to  draw the Carter-Penrose 
conformal diagram much more easily than the case when one employs 
the usual Boyer-Lindquist-type time coordinate. 
Therefore when investigating various physical contents of the spinning 
BTZ black hole, the two time coordinates $t$ and $\tilde{t}$ appear
to play mutually complementary roles.

\vspace{2cm}

{\bf \large References}

\begin{description}

\item {[1]} M. ~Ba$\tilde{\rm{n}}$ados, C. ~Teitelboim, and J. ~Zanelli, 
Phys. Rev. 
Lett. {\bf 69}, 1849 (1992) ;  M. ~Ba$\tilde{\rm{n}}$ados, M. ~Henneaux, 
C. ~Teitelboim, and J. ~Zanelli, Phys. Rev. {\bf D48}, 1506 (1993).
\item {[2]} E. T. ~Newman and A. I. ~Janis, J. Math. Phys. {\bf 6}, 915 
(1965) ;
E. T. ~Newman, E. ~Couch, R. ~Chinnapared, A. ~Exton, A. ~Prakash and
R. ~Torrence, J. Math. Phys. {\bf 6}, 918 (1965).
\item {[3]} R. H. ~Boyer and R. W. ~Lindquist, J. Math. Phys. {\bf 8}, 265 
(1967) ; B. ~Carter, Phys. Rev. {\bf 174}, 1559 (1968).
\item {[4]} R. M. ~Wald, {\it General Relativity} (Univ. of Chicago Press, 
Chicago, 1984).
\item {[5]} J. D. ~Bekenstein, Phys. Rev. {\bf D7}, 2333 (1973) ; {\it ibid},
{\bf D9}, 3292 (1974).
\item {[6]} S. W. ~Hawking, Commun. Math. Phys. {\bf 43}, 199 (1975).
\item {[7]} G. W. ~Gibbons and S. W. ~Hawking, Phys. Rev. {\bf D15}, 2752 
(1977).

\end{description}

\end{document}